\documentstyle[twoside,fleqn,espcrc2,epsf]{article}

\title{QCDSP:  The first 64 nodes}

\author{Robert D. Mawhinney\address{
  Department of Physics, Columbia University, New York, NY 10027,
  USA}\thanks {This
    work was done in collaboration with Dong Chen, Norman
    H. Christ, Chulwoo Jung, Adrian Kaehler, Steve Kasow,
    Yubing Luo, Catalin
    Malureanu, ChengZhong Sui and Pavlos Vranas from Columbia
    University; John Parsons and
    Alan Gara, Columbia University Nevis Laboratory; Tony
    Kennedy and Robert Edwards, SCRI; Greg Kilcup, The
    Ohio State University; Jim Sexton, Trinity College;
    Sten Hansen, Fermilab.  This work was supported in part by the 
    Department of Energy.  Presented at Lattice '96.}
  }
\begin{document}

\begin{abstract}
We present a summary of the progress on QCDSP in the last year.
QCDSP, Quantum Chromodynamics on Digital Signal Processors, is
an inexpensive computer being built at Columbia that can
achieve 0.8 teraflops for three million dollars.
\end{abstract}
       
\def\thepage{CU--TP--777}
\thispagestyle{myheadings}

\maketitle

\section{INTRODUCTION}

QCDSP (Quantum Chromodynamics on Digital Signal Processors) is the name
of the single precision, Digital Signal Processor (DSP) based computer
being built at Columbia.  DSP's are essentially general purpose
microprocessors, whose design has been optimized for inexpensive
floating point power.  The project was begun in April, 1993 as an
inexpensive way to obtain a machine that would give almost teraflop
scale performance for three million dollars.  In particular, for this
price a machine with 16,384 processing nodes, connected as a $16^3
\times 4$ four-dimensional hypercubic array with a nearest neighbor
communications network, would have a peak speed of 0.8 teraflops.

Each processing node of QCDSP contains a 50 Mflop Texas Instruments
TMS320C31 DSP, 2 MBytes of DRAM and a custom Application Specific
Integrated Circuit (ASIC), called the Node Gate Array (NGA).  The NGA
was designed at Columbia and provides three major functionalities.
First, it contains a programmable fetch-ahead memory buffer to increase
the effective bandwidth of the DRAM for regular patterns of memory
access.  Second, it handles DRAM refresh and single bit error detection
and correction.  Its third major function is to provide a 25 or 50 MHz
(the speed is programmable), bit serial communication link to each of
the eight nearest neighbor processoring nodes.  This functionality of
the NGA is handled by the serial communications unit (SCU).  A fully
assembled daughter board costs about \$120 in quantity.

The initial design for QCDSP was described at Lattice '93 \cite{nhc},
details of the processing nodes, NGA design and features at Lattice '94
\cite{rdm} and final NGA design issues, the networks in QCDSP and some
details of the global architecture at Lattice '95 \cite{vranas}.  Since
Lattice '95, the design for our processing nodes, motherboards,
backplanes and crates has been completed and the design translated into
working hardware, demonstrated at Lattice '96 in the form of a working
64 processor machine.  In this paper, we will describe some of the
milestones in this process, give more details about the hardware
architecture of the full machine and discuss our evolving software
environment.

\section{GLOBAL ARCHITECTURE}

Most of the processing nodes described in the introduction are realized
on a roughly $1.8$" by $2.7$" printed circuit board.
We refer to these as daughter boards.  Our motherboards are roughly
14.5" by 20.5", 10 layer printed circuit
boards.  A motherboard holds 63 daughter boards attached to it through
SIMM connectors.  (These daughter boards can be replaced in under a
minute if one fails.)  A 64th processing node, referred to as node 0, is
soldered directly to the motherboard.  A motherboard contains a $4
\times 4 \times 2 \times 2$ configuration of processors, with the first
dimension periodic on the motherboard.  Eight motherboards are included
in a crate and four crates are stacked into a rack.

In any numerical calculation, all the processing nodes behave
similarly.  Node 0 on each motherboard has capabilities that the
daughter boards do not.  While the daughter boards have 2 MBytes of
local DRAM, node 0 has 8 Mbytes.  Node 0 also has special functionality
that is most apparent at boot time, during hardware testing and while
doing I/O.  It controls the two independent SCSI buses that can be
attached to a motherboard, the connections of the DSP serial network
described below, access to the single PROM on the motherboard and the
electronics driving the off-board physics network connections.

\begin{figure}[htb]
\epsfxsize=\hsize
\epsfbox{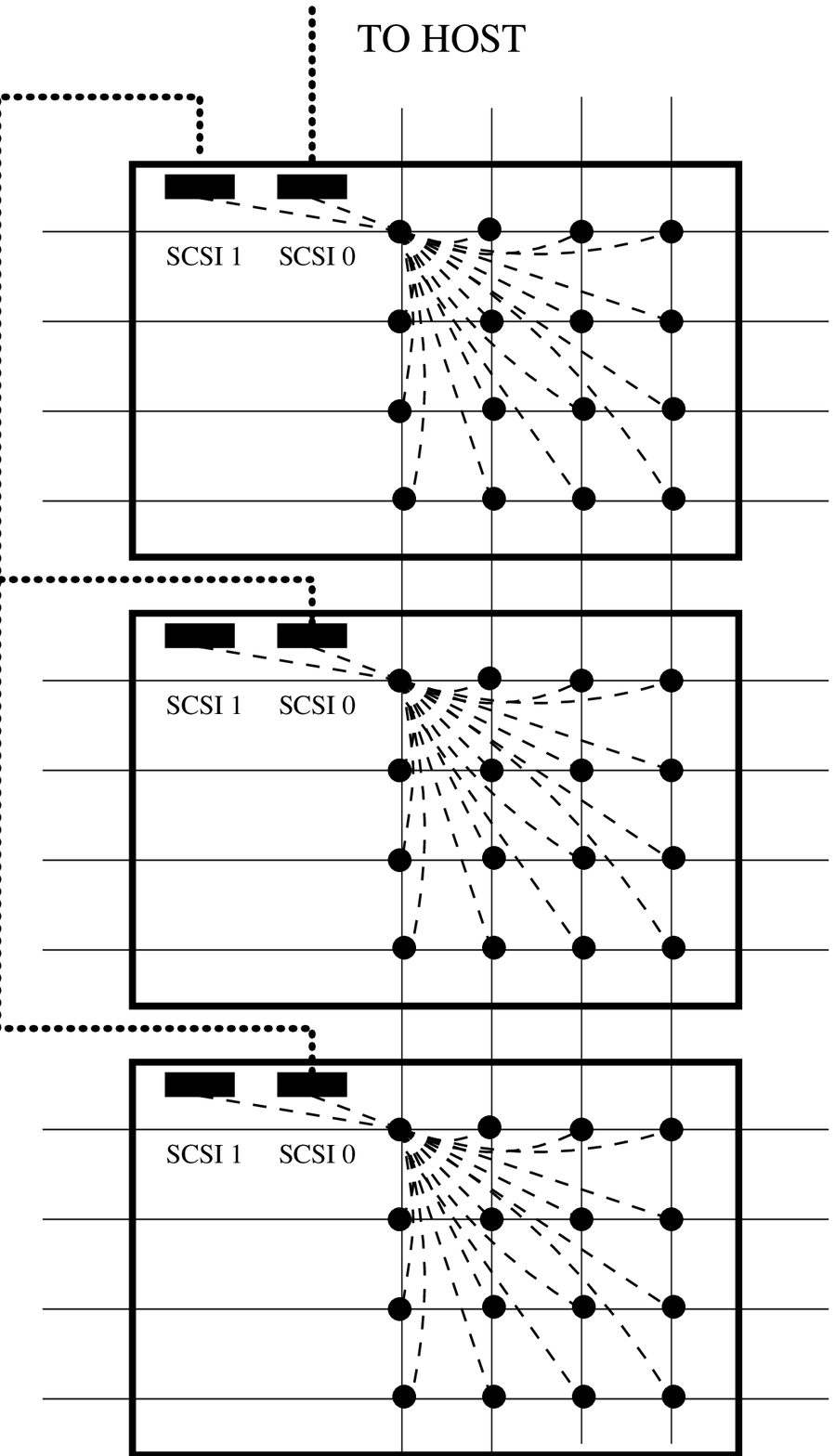}
\caption{A diagram of the various communications paths in QCDSP.
Here the nodes are depicted as located in a two-dimensional
mesh (for ease of exposition) rather than in a four-dimensional
mesh actually implemented.  The filled circles represent
processing nodes, the thin straight lines the four-dimensional
nearest neighbor network, the thin dashed lines the DSP serial
network on the motherboards and the thick dotted lines the
connections in the SCSI tree.  The thick solid lines making
up boxes give the physical boundary of the motherboard.}
\label{fig:networks}
\end{figure}

In addition to the four-dimensional nearest neighbor network described
in the introduction, each motherboard has a DSP serial network
\cite{vranas}.  Figure \ref{fig:networks} shows these two networks,
plus an example of a SCSI tree connection which links motherboards to
each other and the host workstation.  The SCSI network allows the host
workstation to communicate with node 0 on any motherboard.  The DSP
serial network allows node 0 on a motherboard to broadcast to all 63
daughter boards or read and write to one daughter board at a time.
The SCSI tree plus DSP serial network provide a reliable, relatively
slow network for boot-time I/O and hardware verification.

\pagenumbering{arabic}
\addtocounter{page}{1}

\section{HARDWARE COMPONENTS}

\subsection{PROCESSING NODES}

There are seven surface mounted integrated circuit chips on a daughter
board; one DSP, one NGA and 5 DRAM chips.  The NGA can drive two
different types of physical chips (either 512k by 8 bit DRAMs or 256k
by 16 bit DRAMs) to help insulate us from changes in the available
configurations of memory.  From a programmer's perspective, the type of
physical memory used is irrelevant.

We took delivery of the first 10 samples of the gate array in late July
1995.  Within a week, we had assembled 4 working daughter boards.  By
far the largest problem we encountered was doing the surface mount
soldering on the fine pitch (pin spacing of 0.020") NGA.  In March
1996, we had 131 daughter boards commercially assembled.  Using test
software we developed, the 10 daughter boards which had assembly
problems were fixed in about 45 minutes by the assemblers.
This example
lends credence to the idea of having thousands of daughter boards
commercially assembled with few failures.

\subsection{MOTHERBOARDS}

The initial sketch of a motherboard in \cite{nhc} is surprisingly
similar to the final result; surface mounting node 0 directly to the
board is the one major change.  During the last year, the motherboard
circuitry was finalized and the resulting 10 layer printed circuit
board was laid out at Columbia.  The first two motherboards arrived in
December and were hand assembled by us.  By mid-January, all the
peripheral systems associated with node zero were working.  The
remainder of the systems were debugged by mid-March, after the arrival
of enough daughter boards to fully populate the motherboard.
The motherboard displayed at Lattice '96 was the first one
completely assembled.

\subsection{BACKPLANES AND CRATES}

During the spring and summer of 1995, the specification of the
backplane electronics was finalized.  The backplane design
includes an equal-time, 50 MHz clock fanout to
all the crates in the machine.  This is important since nearest
neighbors in the four dimensional network will generally be in
different crates.  An equal time reset signal is also distributed.
Even though the machine is self-synchronizing on nearest neighbor
links, it can be useful for debugging purposes to have all processors
come up from reset synchronously.  Another vital reason for a
synchronous reset is so all nodes will agree on the correct phase for a
25 MHz signal derived from the 50 Mhz clock.  When the SCUs are running
at 25 MHz, all nodes must agree on this signal.

The backplane also handles the distribution of three global interrupts
linking all nodes.  One interrupt gives a global synchronization
signal, the second a non-recoverable error signal and the third as a
recoverable error signal.

The backplane circuit boards were laid out and assembled by
Bustronics.  The crates were designed and built by Elma Corporation,
with the first single crate arriving in December 1995.  The racks of
four crates are under development currently.  Each crate requires a
separate 20 amp, 220 volt circuit and a single crate is cooled by a
tray of muffin fans.  The four crates that are stacked into a rack will
have water-cooled heat exchangers between each crate.

For a given physical configuration of the crates, the four
dimensional configuration of QCDSP is
determined by external cables connected to the backplane.
In particular, the cabling can produce a machine of dimension
$4 \times 4i \times 2j \times 2k$ where $i$, $j$ and $k$ are integers.
Two and three dimensional arrays are also possible by only
altering the external cabling.

\section{SOFTWARE}

\subsection{2 NODE PROTOTYPE}

A rudimentary operating system was written for a 2 node machine,
connected to a host SUN computer through a commercial SBUS DSP board.
This operating system provides read, write and execute access to
the prototype.  It has proved very useful for code development
and testing.  The members of the collaboration from SCRI have
ported much of their macro-based QCD code to this 2 node
prototype, providing additional testing of the hardware.
A quenched update code has been written at Columbia and tested
on the prototype.

\subsection{SOFTWARE FOR HARDWARE TESTING}

Substantial effort has gone into this area.  After assembling the first
daughter boards, they were subjected to a variety of tests for
reliability, correct numerical results and power consumption.
Concurrent with the assembly of the first two motherboards, software
was written to test each system as it was completed.  This software
will become part of the standard boot-time hardware checking done on
QCDSP.

We have already mentioned the automated testing software used to check
the commercial daughter board assembly.  This was so successful that we
are currently packaging our motherboard hardware test software into a
series of diagnostic and functional tests to be run at the assemblers.
After daughter boards are assembled and tested, they will be inserted
into a motherboard and the entire assembly tested.  We expect this will
result in fully populated and competely functional boards being shipped
to Columbia.

\subsection{QCDSP OPERATING SYSTEM SOFTWARE}

We have completed the most basic level of the QCDSP operating system,
called the boot kernels.  The boot kernels are running after machine
reset and provide a robust I/O path to all nodes in QCDSP.  These
kernels fit into the 2k words of DSP internal memory and utilize only
the DSP serial network and the SCSI tree.  This reliance on a minimal
amount of working hardware allows us to do much hardware debugging
through software.

The boot kernel on node 0 includes a fully functional SCSI driver as
well as control features to switch data to various daughter boards.  We
have standardized a recursive routing protocol which can handle an
arbitrary number of layers in the SCSI tree.  We have used the boot
kernels to communicate with a second motherboard and a disk from the
first motherboard.

On our host SUN, we have an X-windows interface to the machine, which
calls a custom SCSI driver to handle I/O to QCDSP.  C programs compiled
with the standard Texas Instruments tools can be loaded and run and the
contents of memory on any node of QCDSP returned to the screen.  The
boot kernels, coupled with the interface, provide full read, write and
execute capability on any node or collection of nodes.  At SCRI, a
command line interface has also been written.

The run kernels are currently being developed.  They will
include all the features of the boot kernels, but will be larger
since DRAM will be available when they are started.  They will
allow for I/O (such as printf) to the host, access to the disk
system, interrupt handlers for machine errors and simple
access to various hardware components.

\subsection{PHYSICS SOFTWARE}

The optimized versions of the staggered and Wilson conjugate gradient
codes written when we were designing the NGA have run on single node, 2
node and 64 node physical machines. In the last year, a quenched
evolution has been written at Columbia.  A dynamical fermion update
code and gauge fixing code are under current development.  In addition,
the macro-based QCD code used by collaborators at SCRI has been ported
to QCDSP.

We have purchased a commercial C++ compiler specifically for the Texas
Instruments DSP from Tartan, Inc.  This compiler includes double
precision arithmetic libraries.  These will run quite slowly since the
hardware is single precision, but will allow us to easily check code
for stability under increased precision.  We hope to develop C++
classes which will hide the parallel nature of the machine from the
user.  These should be useful in the parts of programs that are not
floating-point intensive.

\section{CONCLUSIONS}

In the last year we have built, or had built, a few of each
hardware component of QCDSP.  Shortly after Lattice '96, we
successfully ran physics programs on a 2 motherboard, 128 node
machine.  A fully populated 512 node, 25 Gflop machine should
be finished by the end of September.  All the parts are on
hand for a 2048 node, 100 Gflop machine, which should be completed by
the end of November.  At that time we expect to begin to purchase
components and have them assembled for a 400 Gflop machine (8192 nodes)
to be completed by the spring of 1997.

\end{document}